\documentclass[aps,prl,showpacs,twocolumn]{revtex4}
\usepackage{amsfonts}
\usepackage{amssymb}
\usepackage{amsmath}
\usepackage{graphicx}
\usepackage{epsfig}

\begin{document}

\title{Photon-like flying qubit in the coupled cavity array}
\author{Ying Li$^{1}$}
\author{M.X. Huo$^{1}$}
\author{Z. Song$^{1}$}
\email{songtc@nankai.edu.cn}
\author{C.P. Sun$^{2}$}
\email{suncp@itp.ac.cn}
\homepage{http://www.itp.ac.cn/~suncp}
\affiliation{$^{1}$Department of Physics, Nankai University, Tianjin 300071, China}
\affiliation{$^{2}$Institute of Theoretical Physics, Chinese Academy of Sciences,
Beijing, 100080, China}

\begin{abstract}
We propose a feasible scheme to realize a spin network via a coupled cavity
array with the appropriate arrangement of external multi-driving lasers. It
is demonstrated that the linear photon-like dispersion is achievable and
this property opens up the possibility of realizing the pre-engineered spin
network which is beneficial to quantum information processing.
\end{abstract}

 \maketitle

\emph{Introduction.} Performing a perfect quantum state transfer (QST)
between two distant qubits is highly desirable for quantum computer
architecture. A photon-like \textquotedblleft flying
qubit\textquotedblright\ \cite{Cirac97, Tob, YSPRA} in solid state system is
crucial for scalable quantum computation \cite{DiVincenzo-Fort}. The quantum
spin lattice is a paradigm in condensed matter physics. It serves as a
communication channel to link quantum solid state registers without the need
of conversion among different types of qubits since the work of S. Bose \cite%
{Bose-spin chain}. To perform a high-fidelity state transfer in a
quantum spin network the main obstacle is the non-linear
dispersion relation which reduces the fidelity of QST due to the
dispersive effect for a single magnon, an elementary excitation
corresponding to spin wave. It has been found that nonuniform
nearest neighbor (NN) coupling can allow the perfect QST
\cite{Christandl, Yung, Shi05}. Moreover, non-trivial long-range
coupling distribution is also possible to achieve the linear
photon-like dispersion relation \cite{YS-ChinS}. In practice, it
remains the challenge to realize an optimal coupling strength
distributions in quantum device. Recent developments of technology
in coupled cavity arrays offer the ability to experimentally
observe the quantum many body phenomena and design the
controllable quantum devices for quantum information process. The
high Q cavity and strong interaction between the cavity mode and
atoms have been observed experimentally \cite{HQ,SC}. It has been
proposed to realize an
effective quasi-spin model through the exchanging of virtual photons \cite%
{Bose-XY,Plenio-ES}. This opens the possibility for the application of the
theoretical results in practice.

In this letter, we propose a feasible scheme to realize a quantum
channel for a photon-like flying qubit in a coupled cavity array
with each cavity containing a single three-level atom
\cite{Plenio-ES}. It is shown that the energy-band broadening for
photons can induce a long-range interaction between atoms trapped
in different cavities. Moreover, the distribution of the
long-range interaction strength is optically controllable, which
opens a possibility to pre-engineer a standard $XY$ spin model via
tuning the external multi-driving lasers. As an application, using
a simple optimal setup, we demonstrate that the linear photon-like
dispersion relation for a magnon is achievable.


\begin{figure}[tbp]
\includegraphics[bb=27 243 576 576, width=4.0 cm]{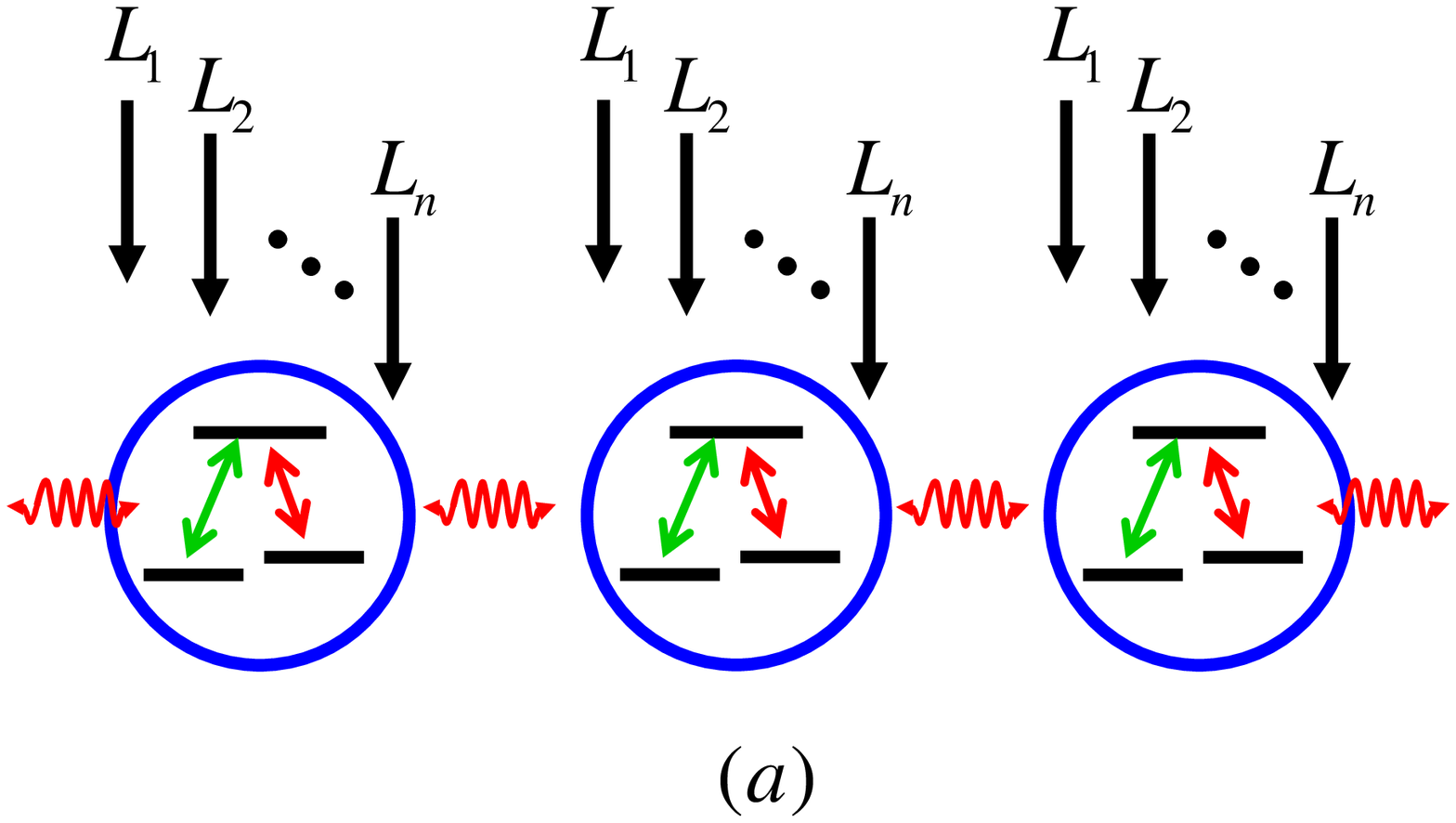} %
\includegraphics[bb=63 177 528 636, width=4.0 cm]{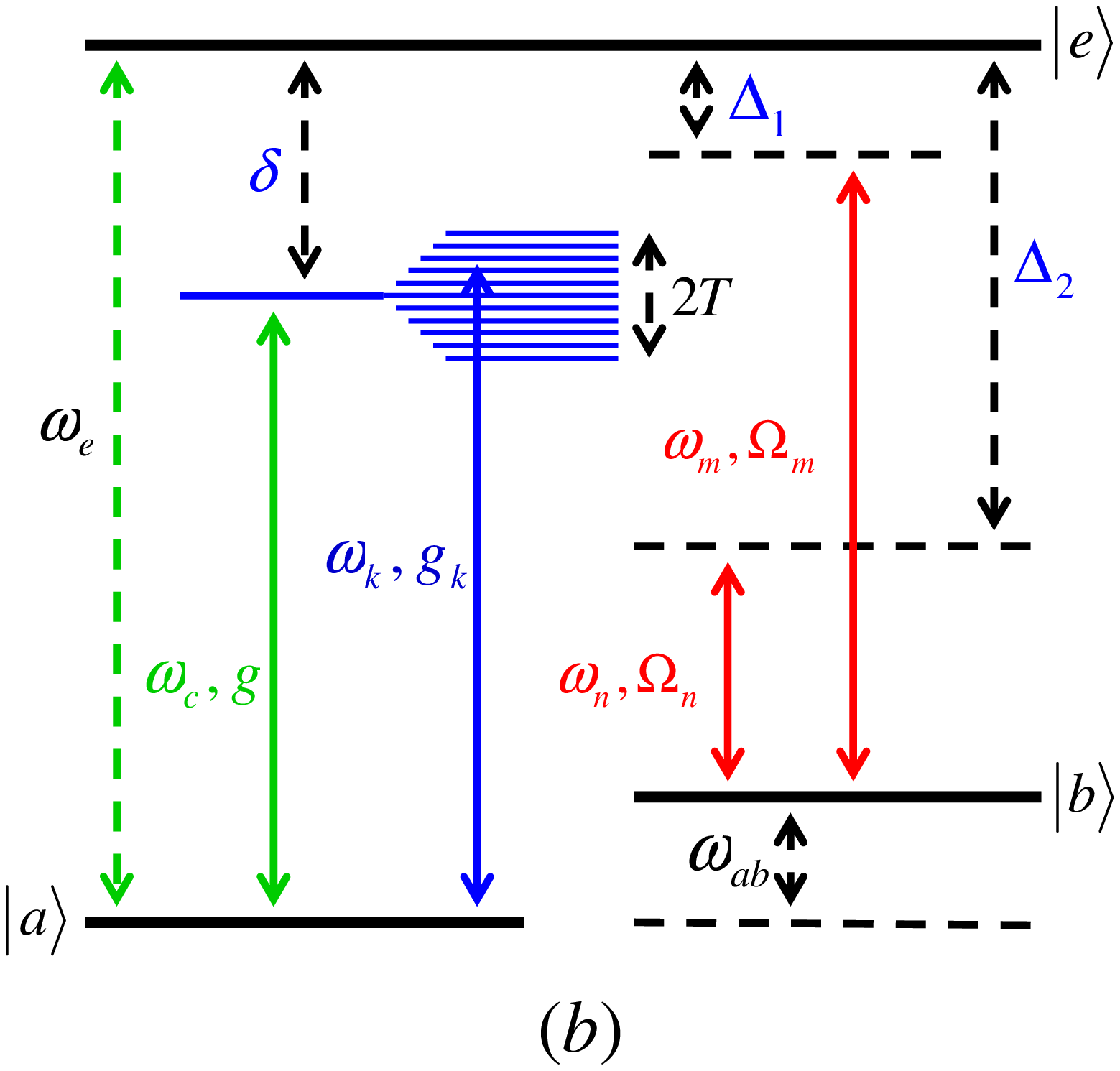}
\caption{\textit{Schematic diagram of }$\Lambda $\textit{-type three-level
atoms in coupled cavities. Each atom interacts with a single mode cavity and
two external driving lasers. Levels }$\left\vert a\right\rangle $\textit{, }$%
\left\vert b\right\rangle $\textit{\ and }$\left\vert e\right\rangle $%
\textit{\ have the energy values }$0$\textit{, }$\protect\omega _{ab}$%
\textit{\ and }$\protect\omega _{e}$\textit{, respectively. The transition }$%
\left\vert a\right\rangle \rightarrow \left\vert e\right\rangle $\textit{\
is driven by photons with the frequency }$\protect\omega _{c}$\textit{\ and
interaction strength }$g$\textit{. }$\protect\omega _{1}$\textit{\ and }$%
\protect\omega _{2}$\textit{\ are the frequencies of the driving lasers with
Rabi frequencies }$\Omega _{1}$\textit{\ and }$\Omega _{2}$\textit{\
respectively. When cavities are coupled with each other, the degenerate
cavity level }$\protect\omega _{c}$\textit{\ becomes an energy band with
bandwidth }$2T$\textit{. Then the coupled cavity array is equivalent to a
multi-mode cavity with frequencies }$\protect\omega _{k}$\textit{.}}
\label{Model}
\end{figure}


\emph{Model setup.} We consider a coupled cavity array with $N$ cavities and
each cavity has a three-level atom. The level structure of the atom trapped
in the $j$th cavity is schemed in Fig. 1(b), where the atom contains two
long-lived states $\left\vert a\right\rangle _{j}$ and $\left\vert
b\right\rangle _{j}$ and an excited state $\left\vert e\right\rangle _{j}$.
The transition between levels $\left\vert a\right\rangle _{j}$ and $%
\left\vert e\right\rangle _{j}$ is coupled to the cavity mode $\omega _{c}$
with the coupling $g$, while the transition between $\left\vert
b\right\rangle _{j}$ and $\left\vert e\right\rangle _{j}$ is driven by $%
n_{L} $ lasers $\left\{ \omega _{n}\right\} $ with the Rabi frequency $%
\left\{ \Omega _{n}\right\} $. The total Hamiltonian is%
\begin{equation}
H=H_{a}+H_{c}+H_{ac}+H_{aL}.  \label{H1}
\end{equation}%
The Hamiltonian $H_{a}$ which describes $\Lambda $ level-structure atoms
reads%
\begin{equation}
H_{a}=\sum_{j}\left( \omega _{e}\left\vert e\right\rangle _{j}\left\langle
e\right\vert +\omega _{ab}\left\vert b\right\rangle _{j}\left\langle
b\right\vert \right) .  \label{Ha}
\end{equation}%
$H_{c}$ describes photons in coupled cavities as%
\begin{equation}
H_{c}=\sum_{j}\omega _{c}a_{j}^{\dag }a_{j}-T\left( a_{j}^{\dag
}a_{j+1}+h.c.\right) ,  \label{Hc}
\end{equation}%
where $\omega _{c}$ is the frequency of photons, $T$ is the tunnelling rate
of photons between two neighboring cavities, and $a_{j}$ ($a_{j}^{\dag }$)
annihilates (creates) a photon in cavity $j$. $H_{ac}$ represents the
coupling between atoms and the cavity mode as
\begin{equation}
H_{ac}=\sum_{j}\left( g\left\vert e\right\rangle _{j}\left\langle
a\right\vert a_{j}+h.c.\right) ,  \label{Hac}
\end{equation}%
while $H_{aL}$ describes the coupling between atoms and $n_{L}$ driving
lasers
\begin{equation}
H_{aL}=\frac{1}{2}\sum_{j}\sum_{n=1}^{n_{L}}\left( \Omega _{n}e^{-i\omega
_{n}t}\left\vert e\right\rangle _{j}\left\langle b\right\vert +h.c.\right) .
\label{HaL}
\end{equation}

Considering the coupled cavity array as an even $N$ site ring and taking the
Fourier transformation
\begin{equation}
\widetilde{a}_{k}^{\dag }=\frac{1}{\sqrt{N}}\sum_{j}e^{ikj}a_{j}^{\dag },
\label{Fourier}
\end{equation}%
the Hamiltonian $H_{c}$ is diagonalized as
\begin{equation}
H_{c}=\sum_{k}\omega _{k}\widetilde{a}_{k}^{\dag }\widetilde{a}%
_{k}=\sum_{k}\left( \omega _{c}-2T\cos k\right) \widetilde{a}_{k}^{\dag }%
\widetilde{a}_{k},  \label{Hck}
\end{equation}%
where $k=2m\pi /N$, $m=0,1,\ldots ,N-1$. Throughout this paper, $k$ denotes
the momentum. Note that such a transformation provides an equivalent view
about the coupled cavity array. Due to the tunnelling of photons between
neighboring cavities, the degenerate cavity level $\omega _{c}$ becomes an
energy band with bandwidth $2T$. Then the coupled cavity array is equivalent
to a multi-mode cavity with frequencies $\omega _{k}$. In this sense, it is
possible to realize a long-range effective interaction between two atoms
placed in distant cavities. We will show that such long-range couplings can
be employed to construct the spin network, which possesses the approximate
linear photon-like dispersion relation for a magnon.

\emph{Adiabatic elimination of atomic excited and photonic states.} We focus
on the case in which the photon excitation in the cavity array is strongly
suppressed. In this case, the atomic states are always changed after
emitting or absorbing a virtual photon. In the following we will
adiabatically eliminate the atomic excited state $\left\vert e\right\rangle $
and the elimination of photonic states. During these procedures, we only
consider the simplest case that each driving laser contributes to the
effective Hamiltonian independently. This scheme requires that the
parameters satisfy the following conditions
\begin{eqnarray}
&\left\vert \delta _{k}\right\vert ,\left\vert \Delta _{n}\right\vert \gg
\left\vert g\right\vert ,\left\vert \Omega _{m}\right\vert ,\left\vert
\Gamma _{m}^{q}\right\vert ,\left\vert \omega _{lm}\right\vert ,&
\label{condition1} \\
&\forall \text{ }k,q\in \lbrack 0,2\pi ),\forall \text{ }n,m,l\in \lbrack
1,n_{L}]&  \notag
\end{eqnarray}%
and%
\begin{eqnarray}
&\left\vert \Gamma _{n}^{k}\right\vert ,\left\vert \omega _{nm}\right\vert
,\left\vert \Gamma _{n}^{k}-\omega _{lm}\right\vert \gg \left\vert \frac{%
g\Omega _{p}}{\Delta _{p}}\right\vert ,\left\vert \frac{\Omega _{p}^{2}}{%
\Delta _{p}}\right\vert ,&  \label{condition2} \\
&\forall \text{ }k\in \lbrack 0,2\pi ),\forall \text{ },n,m,l,p\in \lbrack
1,n_{L}],&  \notag
\end{eqnarray}%
where
\begin{eqnarray}
\delta _{k} &=&\omega _{e}-\omega _{c}+2T\cos k,  \notag \\
\Delta _{n} &=&\omega _{e}-\omega _{ab}-\omega _{n},  \label{parameter} \\
\Gamma _{n}^{k} &=&\delta _{k}-\Delta _{n},  \notag \\
\omega _{nm} &=&\omega _{n}-\omega _{m}.  \notag
\end{eqnarray}%
Now turn to the interaction picture with
\begin{equation}
H_{0}^{e}=H_{a}+H_{c}  \label{He0}
\end{equation}%
and%
\begin{eqnarray}
H_{1}^{e} &=&\sum_{j}\left[ \left\vert e\right\rangle _{j}\left\langle
a\right\vert \sum_{k}\frac{g}{\sqrt{N}}e^{i\left( kj+\delta _{k}t\right) }%
\widetilde{a}_{k}\right.   \label{He1} \\
&&\left. +\frac{1}{2}\sum_{n=1}^{n_{L}}\Omega _{n}e^{i\Delta
_{n}t}\left\vert e\right\rangle _{j}\left\langle b\right\vert +h.c.\right] .
\notag
\end{eqnarray}%
Through adiabatically eliminating the atomic excited state, the effective
Hamiltonian of $H_{1}^{e}$ is written as
\begin{eqnarray}
H_{2}^{e} &=&-iH_{1}^{e}\left( t\right) \int_{-\infty }^{t}dt^{\prime
}H_{1}^{e}\left( t^{\prime }\right)   \label{He2} \\
&=&-\left[ B+\mathfrak{B}\left( t\right) \right] \sum_{j}\sigma _{z}^{\left(
j\right) }  \notag \\
&&-\sum_{j,k}\left[ \mathfrak{g}_{j}\left( k,t\right) \sigma _{+}^{\left(
j\right) }\widetilde{a}_{k}+h.c.\right] ,  \notag
\end{eqnarray}%
where the pseudo spin operators are
\begin{eqnarray}
\sigma _{z}^{\left( j\right) } &=&\left\vert b\right\rangle _{j}\left\langle
b\right\vert -\left\vert a\right\rangle _{j}\left\langle a\right\vert ,
\label{pseudo} \\
\sigma _{+}^{\left( j\right) } &=&\left\vert b\right\rangle _{j}\left\langle
a\right\vert ,  \notag
\end{eqnarray}%
and
\begin{eqnarray}
B &=&\sum_{n}\frac{\left\vert \Omega _{n}\right\vert ^{2}}{8\Delta _{n}},
\notag \\
\mathfrak{B}\left( t\right)  &=&\sum_{m\neq n}\frac{\Omega _{m}^{\ast
}\Omega _{n}}{4\left( \Delta _{n}+\Delta _{m}\right) }e^{i\omega _{mn}t},
\label{BBG} \\
\mathfrak{g}_{j}\left( k,t\right)  &=&\frac{g}{\sqrt{N}}\sum_{n}\frac{\Omega
_{n}^{\ast }}{2\Delta _{n}}e^{i\left( kj+\Gamma _{n}^{k}t\right) }.  \notag
\end{eqnarray}%
Here an irrelevant constant has been dropped. Note that the Hamiltonian (\ref%
{He2}) is equivalent to a JC model which describes an ensemble of atoms
interacting with a multi-mode cavity.

Next the interaction picture is taken as
\begin{equation}
H_{0}^{p}=H_{a}+H_{c}-B\sum_{j}\sigma _{z}^{\left( j\right) }  \label{Hp0}
\end{equation}%
and
\begin{equation}
H_{1}^{p}=H_{2}^{e}+B\sum_{j}\sigma _{z}^{\left( j\right) }.  \label{Hp1}
\end{equation}%
Eliminating the photonic degree of freedom, we have%
\begin{eqnarray}
H_{2}^{p} &=&-iH_{1}^{p}\left( t\right) \int_{-\infty }^{t}dt^{\prime
}H_{1}^{p}\left( t^{\prime }\right)  \label{Hp2} \\
&=&-\sum_{i,j}\sum_{n=1}^{n_{L}}\left\vert \frac{g\Omega _{n}}{2\Delta _{n}}%
\right\vert ^{2}S_{ij}^{\left[ n\right] }\sigma _{+}^{\left( i\right)
}\sigma _{-}^{\left( j\right) },  \notag
\end{eqnarray}%
where
\begin{equation}
S_{ij}^{\left[ n\right] }=\frac{1}{N}\sum_{k}\frac{e^{ik\left( i-j\right) }}{%
\mathcal{D}_{n}-2T\cos k}  \label{Sij}
\end{equation}%
and
\begin{equation}
\mathcal{D}_{n}=\omega _{c}-\omega _{ab}-\omega _{n}.  \label{Dn}
\end{equation}%
Combining $H_{0}^{p}$ and $H_{2}^{p}$, the effective Hamiltonian of atoms is
obtained as
\begin{eqnarray}
H_{eff} &=&-\sum_{i,j}\frac{J_{ij}}{2}\left( \sigma _{+}^{\left( i\right)
}\sigma _{-}^{\left( j\right) }+h.c.\right)  \label{Heff} \\
&&+\sum_{j}\left( \frac{\omega _{ab}}{2}-B\right) \sigma _{z}^{\left(
j\right) },  \notag
\end{eqnarray}%
where
\begin{equation}
J_{ij}=\sum_{n=1}^{n_{L}}\left\vert \frac{g\Omega _{n}}{2\Delta _{n}}%
\right\vert ^{2}S_{ij}^{\left[ n\right] }.  \label{J_ij}
\end{equation}%
It is a standard $XY$ model with the pre-engineered coupling distribution.
The long-range interaction is the result of the photonic energy-band
broadening. It is also indicated that $J_{ij}=J_{ji}=J\left( \left\vert
i-j\right\vert \right) $.

In the narrow band limit $\left\vert \mathcal{D}_{n}\right\vert \gg
\left\vert 2T\right\vert $, Eq. (\ref{Sij}) becomes
\begin{equation}
S_{ij}^{\left[ n\right] }\simeq \delta _{i,j}\frac{1}{\mathcal{D}_{n}}%
+\delta _{\left\vert i-j\right\vert ,1}\frac{T}{\mathcal{D}_{n}^{2}},
\end{equation}%
which is reduced to the result of the spin model with NN couplings obtained
in Ref. \cite{Plenio-ES}. This setup leads to a cosinusoidal dispersion
relation, i.e., a magnon has a quadratic dispersion relation for small
momenta but a linear photon-like dispersion relation for momenta around $%
k=\pm \pi /2$ \cite{YSPRA, ChenB}.

On the other hand, in the limit $N\rightarrow \infty $, we have
\begin{equation}
S_{ij}^{\left[ n\right] }=\sigma _{n}^{i-j+1}\frac{1}{\sqrt{\mathcal{D}%
_{n}^{2}-4T^{2}}}\exp \left( -\frac{\left\vert i-j\right\vert }{\xi _{n}}%
\right)  \label{ASij}
\end{equation}%
with
\begin{equation}
\xi _{n}^{-1}=-\ln \left[ \left\vert \frac{\mathcal{D}_{n}}{2T}\right\vert -%
\sqrt{\left( \frac{\mathcal{D}_{n}}{2T}\right) ^{2}-1}\right] ,  \label{Xi}
\end{equation}%
where $\xi _{n}$ is the characteristic length of the long-range effective
interaction and $\sigma _{n}=$ sign $\left( \mathcal{D}_{n}/T\right) $.

\emph{Designed spin chain.} In quantum information processing, a qubit state
is usually transferred by photons via the fiber. In order to implement the
scalable quantum computation in solid state systems, a spin chain is of
particular interests because it may act as a data bus to link qubits without
the need of conversion among different types of qubits. Then realizing a
flying qubit in a spin chain is significant. The dynamics of the magnon wave
packet has been studied recently \cite{Tob, YSPRA}. The main obstacle to
perform a high-fidelity state transfer in a quantum spin model is the
spreading of the wave packet, which is due to the nonlinear dispersion
relation of the spin wave. A previous study has shown that a long-range $XY$%
-interaction gives a chance to realize the flying qubit with a linear
photon-like dispersion relation \cite{YS-ChinS}.

For a standard $XY$ model,
\begin{equation}
H_{XY}=-\sum_{i,l}\frac{J\left( l\right) }{2}\left( \sigma _{+}^{\left(
i\right) }\sigma _{-}^{\left( i+l\right) }+h.c.\right) ,  \label{Hxy}
\end{equation}%
where $J\left( l\right) =J\left( -l\right) $. The eigenstates in the
subspace with a single spin flipped on a ferromagnetic background are $%
\left\vert k\right\rangle =$ $1/\sqrt{N}\sum_{j=1}^{N}e^{ikj}\left\vert
j\right\rangle $ with $\left\vert j\right\rangle =\sigma _{+}^{\left(
j\right) }\prod\nolimits_{i=1}^{N}\left\vert \downarrow \right\rangle _{i}$.
The dispersion relation for the single magnon is
\begin{equation}
E_{k}=-J\left( 0\right) -\sum_{l>0}2J\left( l\right) \cos kl.  \label{Ek}
\end{equation}%
Note that, in principle, the system with any dispersion relation can be
constructed by an appropriate distribution of $J\left( l\right) $, which can
be realized via the external driving lasers. For a linear photon-like
dispersion curve $\varepsilon _{k}=\left\vert k\right\vert $, the
corresponding Fourier expansion is \cite{YS-ChinS}
\begin{equation}
\varepsilon _{k}=\frac{\pi }{2}-\frac{2}{\pi }\sum_{l=1}^{\infty }\frac{%
1-\left( -1\right) ^{l}}{l^{2}}\cos lk,  \label{E_Fourier}
\end{equation}%
which requires
\begin{equation}
J\left( 0\right) =-\frac{\pi }{2},\text{ }J\left( l\neq 0\right) =\frac{%
1-\left( -1\right) ^{l}}{\pi l^{2}}.  \label{J(0)}
\end{equation}%
On the other hand, we can use $n_{L}/2$ pairs of driving lasers with%
\begin{eqnarray}
\left\vert \frac{\Omega _{2m-1}}{\Delta _{2m-1}}\right\vert &=&\left\vert
\frac{\Omega _{2m}}{\Delta _{2m}}\right\vert =G_{2m};  \label{nL} \\
\mathcal{D}_{2m-1} &=&-\mathcal{D}_{2m},\text{ }m\in \lbrack 1,n_{L}/2]
\notag
\end{eqnarray}%
to simulate the coupling as%
\begin{equation}
J_{ij}=\frac{g^{2}}{4}\sum_{m=1}^{n_{L}/2}G_{2m}^{2}\left\vert S_{ij}^{\left[
2m\right] }\right\vert \left[ 1-\left( -1\right) ^{i-j}\right] .
\label{J_ij2}
\end{equation}%
It is indicated that the photon-like dispersion relation is probably
achievable in a system with an optimal arrangement of $n_{L}/2$ pairs of
driving lasers. In practice, it is impossible to set up many external
lasers. However, a straightforward calculation shows that a feasible setup
with several deriving lasers can work efficiently. Consider the optimal
parameters in two simple cases with $n_{L}=2$ and $4$ where for $n_{L}=2$, $%
\mathcal{D}_{2}=10T/3$ and for $n_{L}=4$, $\mathcal{D}_{2}=20T$, $\mathcal{D}%
_{4}=34T/15$, and $\left\vert \Omega _{2}\Delta _{4}/\Omega _{4}\Delta
_{2}\right\vert =6\sqrt{14}$. Other parameters are set to satisfy the
condition (\ref{Dn}). In Table 1, the distribution of the coupling constant $%
J\left( l=\left\vert i-j\right\vert \right) =J_{ij}$ is displayed in unit $%
J(1)$ from (\ref{J_ij2}) and from the ideal case (\ref{J(0)}). The
corresponding dispersion curves and group velocities are plotted in Fig. 2
and compared with those of a photon-like cosinusoidal dispersion. The
obtained results indicate that the ideal distribution of the coupling
constant (\ref{J(0)}) and the photon-like dispersion relation are achievable
approximately using our simple and optimal setups with $n_{L}=2$ and $4$.

To prove our scheme, the time evolution of a single qubit state is
considered in a ring with $N=40$ and with the optimal coupling distribution
for $n_{L}=4$ as listed in Table 1. The initial state is $\left\vert \psi
\left( N_{0}\right) \right\rangle $ with $N_{0}=10$. Driven by the
Hamiltonian with the ideal coupling distribution (\ref{J(0)}), the state
separates into two local moving wave packets, and they meet at site $%
N_{0}+N/2$ to form a single qubit state $\left\vert \psi \left(
N_{0}+N/2\right) \right\rangle =\left\vert \Psi \left( 30\right)
\right\rangle $ after a period of time $\tau =N\pi /8$ \cite{YS-ChinS}. The
profile of the time evolution is plotted in Fig. 3 numerically. It exhibits
a \textquotedblleft whispering gallery\textquotedblright\ behavior, which is
crucial for the QST and entanglement creation.

\begin{center}
\begin{tabular}{cccccccc}
\hline\hline
& $l$ & $1$ & $3$ & $5$ & $7$ & $9$ & $11$ \\ \hline
& $n_{L}=2$ & $\frac{1}{1^{2}}$ & $\frac{1}{3.0^{2}}$ & $\frac{1}{9.0^{2}}$
& $\frac{1}{27.0^{2}}$ & $\frac{1}{81.0^{2}}$ & $\frac{1}{243.0^{2}}$ \\
$\frac{J\left( l\right) }{J\left( 1\right) }$ & $n_{L}=4$ & $\frac{1}{1^{2}}$
& $\frac{1}{3.0^{2}}$ & $\frac{1}{5.0^{2}}$ & $\frac{1}{8.4^{2}}$ & $\frac{1%
}{14.0^{2}}$ & $\frac{1}{23.2^{2}}$ \\
& ideal & $\frac{1}{1^{2}}$ & $\frac{1}{3^{2}}$ & $\frac{1}{5^{2}}$ & $\frac{%
1}{7^{2}}$ & $\frac{1}{9^{2}}$ & $\frac{1}{11^{2}}$ \\ \hline
\end{tabular}
\end{center}

\textit{Table 1. The distribution of the coupling constant in setups with
optimal parameters for }$n_{L}=2,$\textit{\ }$4$\textit{\ and in the ideal
case (\ref{J(0)}). It indicates that the ideal coupling distribution is
achievable approximately via our scheme.}


\begin{figure}[tbp]
\includegraphics[bb=48 254 542 646, width=7.0 cm]{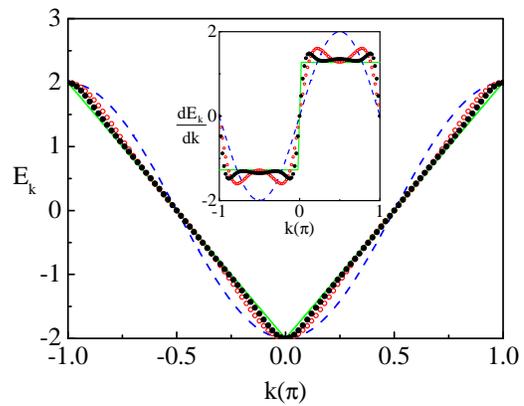}
\caption{\textit{Dispersions with different coupling distributions for
(empty circle) }$n_{L}=2$\textit{, (solid circle) }$n_{L}=4$\textit{,\ and
(solid line) the ideal case, as given in Table 1. The dash line corresponds
to the case with NN couplings. The inset shows the derivatives (group
velocities) of dispersions.}}
\label{SpectrumV}
\end{figure}



\begin{figure}[tbp]
\includegraphics[bb=22 310 494 747, width=7.0 cm]{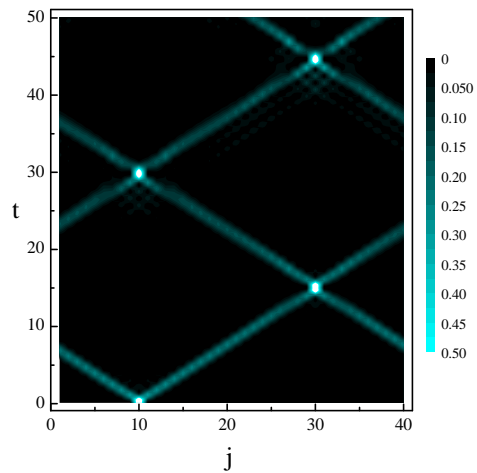}
\caption{\textit{Profile of the time evolution of a single-qubit state in a
ring with }$N=40$\textit{\ and with the optimal coupling distribution for }$%
n_{L}=4$\textit{, as\ listed in Table 1. The initial state is set as }$%
\left\vert \protect\psi \left( N_{0}\right) \right\rangle $\textit{\ with }$%
N_{0}=10$\textit{. The exhibited \textquotedblleft whispering
gallery\textquotedblright\ behavior justifies our scheme.}}
\label{WP}
\end{figure}


\emph{Summary.} In summary, we have shown that the feasible scheme
to realize a quantum channel for a photon-like flying qubitis
achieveable in a coupled cavity array with each cavity containing
a single three-level atom. The underlying physics is best
understood as that the energy-band broadening for photons can
induce a long-range interaction between atoms trapped in different
cavities. We also show that a simple optimal setup with several
external driving lasers can realize the linear photon-like
dispersion relation for a magnon. We have demonstrated that this
property opens up the possibility of realizing the pre-engineered
spin network which is beneficial to quantum information
processing.

This work was supported by the NSFC with Grant Nos. 90203018,
10474104, 60433050, 10547101, 10604002, and 10704023, NFRPC with
Grant Nos. 2001CB309310, 2005CB724508, and 2006CB921205.


\begin{thebibliography}{99}
\bibitem{Cirac97} J.I. Cirac \textit{et al.}, Phys. Rev. Lett. \textbf{78},
3221 (1997).

\bibitem{Tob} T.J. Osborne and N. Linden, Phys. Rev. A \textbf{69}, 052315
(2004).

\bibitem{YSPRA} S. Yang, Z. Song, and C.P. Sun, Phys. Rev. A \textbf{73},
022317 (2006).

\bibitem{DiVincenzo-Fort} D.P. DiVincenzo, Fortschr. Phys. \textbf{48}, 9
(2000).

\bibitem{Bose-spin chain} S. Bose, Phys. Rev. Lett. \textbf{91}, 207901
(2003).

\bibitem{Christandl} M. Christandl et al., Phys. Rev. Lett. 92, 187902 2004
; Phys. Rev. A \textbf{71}, 032312 (2005).

\bibitem{Yung} M.H. Yung and S. Bose, Phys. Rev. A 71, 032310 (2005).

\bibitem{Shi05} T. Shi \textit{et al.}, Phys. Rev. A \textbf{71}, 032309
(2005).

\bibitem{YS-ChinS} S. Yang, Z. Song, and C.P. Sun, Science in China \textbf{%
51}, 45 (2008).

\bibitem{HQ} D.K. Armani \textit{et al.}, Nature \textbf{421}, 925 (2003).

\bibitem{SC} T. Aoki \textit{et al.}, Nature \textbf{443}, 671 (2006).

\bibitem{Plenio-ES} M.J. Hartmann, F.G.S.L. Brand\~{a}o, and M.B. Plenio,
Phys. Rev. Lett. \textbf{99}, 160501 (2007).

\bibitem{Bose-XY} D.G. Angelakis, M.F. Santos, and S. Bose, Phys. Rev. A
(Rap. Com.) \textbf{76}, 031805 (2007).

\bibitem{ChenB} B. Chen, Z. Song, and C.P. Sun, Phys. Rev. A \textbf{75},
012113 (2007).
\end{thebibliography}
\end{document}